\documentstyle[aps,preprint]{revtex}

\tighten
\draft
\begin{document}

\preprint{RUB-TPII-5/98,~~~  TPJU-8/98}
\title{SPECTRUM OF THE ODDERON CHARGE FOR
ARBITRARY CONFORMAL WEIGHTS}
\author{Micha{\l} Prasza{\l}owicz\thanks{%
A.v. Humboldt fellow on leave of absence from the Institute of Physics,
Jagellonian University, Cracow, Poland}$^{\dagger}$
and Andrzej Rostworowski$^{\dagger}$}
\address{~~\\
$^{*}$Institute for Theoretical Physics II,\\
Ruhr-University Bochum, \\
D--44780 Bochum, Germany \\
~~~\\
$^{\dagger}$Institute of Physics, \\
Jagellonian University, \\
ul. Reymonta 4,\\
PL--30-059 Krak{\'o}w, Poland. \\
~~}
\date{\today}
\maketitle

\begin{abstract}
The odderon equation is studied in terms of the variable suggested by the
modular invariance of the 3 Reggeon system. Odderon charge is identified
with the cross-product of three conformal spins. A complete set of commuting
operators: $\hat{h}^2$ and $\hat{q}$ is diagonalized and quantization
conditions for eigenvalues of the odderon charge $\hat{q}$ are solved for
arbitrary conformal weigth $h$.
\end{abstract}


\input{epsf.tex}

%%%%%%%%%%%%%%%%%%TEXT

In 1980, following Lipatov and collaborators \cite{KLF,BL}, the
integral equation for the exchange of 3 or more reggeized gluons has
been formulated \cite{Bart80}\nocite{KwPra} --\cite{Jar}. This equation
describes both the leading contribution to the odderon (odd $C$ parity)
exchange, as well as the unitarity correction to the Pomeron. Although
originally the odderon equation has been written in the momentum space,
it turns out, that it is convenient to rewrite it in the 2 dimensional
configuration space of the impact parameters $b_i=(x_i,y_i)$, where
$i=1,2$ or 3 (or $n$ for amplitudes with more Reggeon exchange)
\cite{Lipatov}. It has been observed that the odderon problem is
equivalent to the system of 3 conformal spins \cite
{Lipsolv}\nocite{FadKor} --\cite{KorBethe} associated with each Reggeon
$i$:
\begin{equation}
S_{+}^{(i)}=z_i^2\partial _i,\;\;\;\;\ S_3^{(i)}=z_i\partial _i
\;\;\;{\rm and}\;\;\;
S_{-}^{(i)}=-\partial _i.  \label{confSpins}
\end{equation}
Here $z_i=x_i+iy_i$. The odderon intercept
\begin{equation}
\alpha =1-\frac{N_{{\rm c}}\alpha _{{\rm s}}}\pi E,
\end{equation}
is related to the eigenvalue $E$ of the interaction Hamiltonian of the
conformal spins:
\begin{equation}
{\cal H}={\rm const.}\sum\limits_{i>j}^3\left( H(z_i,z_j)+H(\overline{z}_i,
\overline{z}_j)\right) .  \label{ham}
\end{equation}
Equation (\ref{ham}) exhibits conformal separability into holomorphic and
antiholomorphic parts, the latter depending only on
$\overline{z}_i=x_i-iy_i$
. Therefore $E=\epsilon +\overline{\epsilon }$ and the wave function is
given as a bilinear form
$\Phi (z,\overline{z})=\overline{\Psi }(\overline{z})\times \Psi (z)$.
There are two conditions for the total wave function --
1) $\Phi (z,\overline{z})$ has to be single-valued and 2) normalizable --
which determine the spectrum of $E$.

There are two scalars which can be constructed from three spins:
\begin{equation}
\hat{h}^2=-\left( \vec{S}^{(1)}+\vec{S}^{(2)}+\vec{S}^{(3)}\right) ^2\;\;\;%
{\rm and}\;\;\;\hat{q}=-\;\vec{S}^{(1)}\cdot \left( \vec{S}^{(2)}\times \vec{%
S}^{(3)}\right) .  \label{q2q3}
\end{equation}
It has been shown in Refs.\cite{Lipsolv}\nocite{FadKor} --\cite{KorBethe}
that the Casimir operator $\hat{h}^2$ (denoted in the literature as $\hat{q}%
_2$) and $\hat{q}$ (denoted often as $\hat{q}_3$) can be simultaneously
diagonalized. In that sense the model is fully integrable. Eigenvalues $h$
of $\hat{h}^2$ (see Eq.(\ref{q2Psi}))are called conformal weights. In the
following, the operator $\hat{q}$ will be called {\em odderon charge}.

There have been many attempts to find either directly values of $E$ \cite
{GLN}\nocite{ArmBraun,bootstrap} --\cite{Braun98} or spectrum of $\hat{q}$
\cite{Korch,JaWoOdd}. In the recent paper Janik and Wosiek \cite{JaWoOddsol}
calculated the spectrum of $\hat{q}$ for $h=1/2$, which corresponds to the
lowest representation of the SL(2,$C$) group, and found the odderon
intercept with $E=0.24717$.

In the present paper we shall concentrate on calculating the spectrum of the
odderon charge for arbitrary $h$. We shall first consider the holomorphic
sector only; however the same arguments apply to the antiholomorphic sector
as well.

Following Lipatov \cite{Lipatov} we shall use conformal Ansatz for $\Psi $:
\begin{equation}
\Psi (z_1,z_2,z_3)=z^{h/3}\;\psi (x),  \label{Psi}
\end{equation}
where
\begin{equation}
z=\frac{(z_1-z_2)(z_1-z_3)(z_2-z_3)}{%
(z_1-z_0)^2(z_2-z_0)^2(z_3-z_0)^2} \;, \;\;\;\;
x=\frac{(z_1-z_3)(z_3-z_0)}{(z_1-z_0)(z_3-z_2)}  \label{x}
\end{equation}
and $z_0$ is a reference point. A remarkable feature
of Ansatz (\ref{Psi}) is, that $\hat{h}^2$ is automatically diagonal:
\begin{equation}
\hat{h}^2\,\Psi (z_1,z_2,z_3)=-h(h-1)\,\Psi (z_1,z_2,z_3).  \label{q2Psi}
\end{equation}
In the representation (\ref{Psi}) the eigenvalue equation for $q$ takes the
following form:
\begin{eqnarray}
i\hat{q}\,\psi (x) &=&-\left( \frac h3\right) ^2\left( \frac h3-1\right)
\frac{(x-2)(x+1)(2x-1)}{x(x-1)}\,\psi (x)  \nonumber \\
&&-\left[ 2x(x-1)-\frac h3(h-1)\left( x^2-x+1\right) \right] \,\psi ^{\prime
}(x)  \nonumber \\
&&-2x(x-1)(2x-1)\,\psi ^{\prime \prime }(x)\;\;-x^2(x-1)^2\,\psi ^{\prime
\prime \prime }(x).  \label{q3x}
\end{eqnarray}

This equation has been recently studied by Janik and Wosiek in Ref.\cite
{JaWoOddsol}. It has 3 regular singular points: at $x=0$, 1 and $\infty $.
The solutions of the indicial equation around 0 read: $s_0=-h/3$, $%
s_0=-(h-3)/3$ and $s_0=2h/3$. This means that in the vicinity of 0 there are
two power series solutions and a logarithmic one. The same holds for the
solutions around $1$ and $\infty $. The authors of Ref.\cite{JaWoOddsol}
formulated quantization conditions for $q$ by imposing singlevalueness
constraints on the whole wave function $\Phi (z,\overline{z})$, and solved
them for $h=1/2$. Discrete, symmetrically distributed values of $q$ have
been found on the imaginary, as well as on the real axis in the complex $q$
plane. However, only the imaginary values of $q$ are relevant for the
odderon problem; real $q$'s correspond to the wave function which is
antisymmetric if the two neighboring Reggeons are exchanged, whereas the
odderon wave function should be symmetric under such transformations \cite
{JaWoOdd}.

Let us rewrite Eq.(\ref{q3x}) in terms of a new variable:
\begin{equation}
\xi =i\;\frac 1{3\sqrt{3}}\;\frac{(x-2)(x+1)(2x-1)}{x(x-1)}\;,  \label{xi}
\end{equation}
suggested in the Ref.\cite{Janik}, where modular invariance of the odderon
equation has been discussed. This mapping sends all singular points of Eq.(%
\ref{q3x}), {\em i.e.} $x=0$, 1 and $\infty $ to infinity. The advantage of
using variable $\xi $ instead of $x$ consists in the symmetry properties of $%
\xi $ under the cyclic permutations of the three reggeized gluons, which
correspond to:
\begin{equation}
x\rightarrow 1-\frac 1x,~~~{\rm or}~~~~x\rightarrow \frac 1{1-x}.
\label{perms}
\end{equation}
Under transformations (\ref{perms}) $\xi $ remains unchanged.
%The non-cyclic permutation corresponds to:
%\begin{equation}
%x\rightarrow \frac x{x-1}\ \;\text{or \ }\xi\rightarrow -\xi\;.
%\label{oddperms}
%\end{equation}

Equation (\ref{q3x}) takes the following form in terms of $\xi$:
\begin{eqnarray}
\left[ \frac 12(\xi ^2-1)^2\frac{d^3}{d\xi ^3}+2\xi (\xi ^2-1)\frac{d^2}{%
d\xi ^2}+\left( \frac 49-\frac{(h+2)(h-3)}6(\xi ^2-1)\right) \frac d{d\xi }%
\right. &&  \nonumber \\
\left. +\frac{h^2(h-3)}{27}\xi +\frac{q}{3\sqrt{3}}\right] \psi (\xi )
&=&0\;.  \label{q3xi}
\end{eqnarray}

Our strategy consists in applying the method of Ref.\cite{JaWoOddsol} to Eq.(%
\ref{q3xi}) rather than to (\ref{q3x}). The advantage of using (\ref{q3xi})
is twofold: 1) because of the symmetry properties (\ref{perms}) the
quantization condition takes a simpler form than in the case of Eq.(\ref{q3x}%
), 2) since Eq.(\ref{q3xi}) is less singular than Eq.(\ref{q3x}) the
solutions of the indicial equation do not depend on $h$. Because of the
latter it is easy to find spectrum of $\hat{q}$ for an arbitrary $h$.

Equation (\ref{q3xi}) has 3 regular singular points in $\xi = \pm 1$ and in
infinity. In what follows we shall consider only solutions around $\pm1$:
\begin{equation}
u_s^{(1)}(\xi ;q)=(1-\xi )^s\sum\limits_{n=0}^\infty u_n^{(1)}(\xi -1)^n\;,
\;\;\;\;\; u_s^{(-1)}(\xi ;q)=(\xi +1)^s\sum\limits_{n=0}^\infty
u_n^{(-1)}(\xi +1)^n\;.  \label{sols}
\end{equation}
The phases of the two solutions are chosen in such a way, that they are real
for real $-1<\xi <1$ and real $q$. The indicial equation for $s$ has the
following solutions:
\begin{equation}
s_1=\frac 23,~~~~s_2=\frac 13~~~{\rm and}~~~s_3=0\;.  \label{roots}
\end{equation}
Since none of the 3 roots of Eq.(\ref{roots}) differ from the other 2 by an
integer we do not have to construct the logarithmic solution. Introducing
notation:
\begin{equation}
\beta _h=\frac{(h+2)(h-3)}6,~~~\gamma _h=\frac{h(h-1)}6,~~~\rho _h=\frac{%
h^2(h-3)}{27},~~~\tilde{q}=\frac{q}{3\sqrt{3}}  \label{defs}
\end{equation}
we can write the recurenence formula:
\begin{eqnarray}
u_0^{(\pm 1)} &=&1\;,  \nonumber \\
&&  \nonumber \\
u_1^{(\pm 1)} &=&\mp \;\frac{2s(s^2-1-\beta _h)+\rho _h\pm \tilde{q}} {%
2(1+s)\left[ s(1+s)+\frac 29\right] }\;,  \nonumber \\
&&  \nonumber \\
u_{n+2}^{(\pm 1)} &=&\mp \;\frac{2(n+1+s)\left[ (n+s)(n+2+s)-\beta _h\right]
+\rho _h\pm \tilde{q}}{2(n+2+s)\left[ (n+2+s)(n+1+s)+\frac 29\right] }%
\;u_{n+1}^{(\pm 1)}  \nonumber \\
&&  \nonumber \\
&&-\frac{(n+s)\left[ (n-1+s)(n+2+s)-2\beta _h\right] +2\rho _h}{%
4(n+2+s)\left[ (n+2+s)(n+1+s)+\frac 29\right] }\;u_n^{(\pm 1)}\;.
\label{rec1}
\end{eqnarray}

These series are convergent in circles of radius 2. Analogously one can
define solutions in the antiholomorphic sector which in the following will
be denoted as $v_s^{(\pm 1)}(\overline{\xi };\overline{q})$ . The three
solutions corresponding to 3 different $s_i$ values form a vector:
\[
\vec{u}^{(\pm 1)}(\xi ;q)=\left[
\begin{array}{c}
u_{s_1}^{(\pm 1)}(\xi ;q) \\
u_{s_2}^{(\pm 1)}(\xi ;q) \\
u_{s_3}^{(\pm 1)}(\xi ;q)
\end{array}
\right] \, .
\]

The analytical continuation matrix $\Gamma $ is defined in the intersection
of the two convergence circles:
\begin{equation}
\vec{u}^{(-1)}(\xi ;q)=\Gamma (q)\;\vec{u}^{(1)}(\xi ;q) \; .  \label{ancont}
\end{equation}
In order to calculate $\Gamma $ we construct a Wro\'{n}skian:
\begin{equation}
W=\left|
\begin{array}{ccc}
u_1^{(1)}(\xi ) & u_2^{(1)}(\xi ) & u_3^{(1)}(\xi ) \\
{u^{\prime }}_1^{(1)}(\xi ) & {u^{\prime }}_2^{(1)}(\xi ) & {u^{\prime }}%
_3^{(1)}(\xi ) \\
{u^{\prime \prime }}_1^{(1)}(\xi ) & {u^{\prime \prime }}_2^{(1)}(\xi ) & {%
u^{\prime \prime }}_3^{(1)}(\xi )
\end{array}
\right| \;.  \label{u1Wr}
\end{equation}
Next we construct determinants $W_{ij}$, which are obtained from $W$ by
replacing $j$-th column by the $i$-th solution around $-1$. Then:
\begin{equation}
\Gamma _{ij}=\frac{W_{ij}}W\;.
\end{equation}

Matrix $\Gamma (q)$ does not depend on $\xi $, but only on $q$ and also on $%
h $. We choose to calculate it at $\xi =0$. Repeating the same steps in the
antiholomorphic sector
%, where the fundamental solutions in variable $\overline{\xi }$ are 
%denoted by $v_{s_i}^{(\pm 1)}(\overline{\xi };\overline{q})$ , 
one constructs $\overline{\Gamma }(\overline{q})$ where:
\begin{equation}
\overline{q}=-q^{\star }\;.  \label{qbar}
\end{equation}
Here $\overline{q}$ denotes the odderon charge in the antiholomorphic
sector, whereas the {\em star} over $q$ denotes complex conjugation. In
principle two possible choices for $\overline{q}$ , namely with + and $-$
signs should be considered. This follows from the fact that both $\epsilon $
and $\overline{\epsilon }$ are symmetric functions of $q$ (or $\overline{q}$%
) \cite{KorBethe}. However, only the choice of Eq.(\ref{qbar}) leads to the
non-zero solutions of the quantization conditions\footnote{%
Note, that because of the factor of $i$ in the definition of $\xi $ (\ref{xi}%
), our sign for $\overline{q}$ is different than the one in Ref.\cite
{JaWoOddsol}.}.

The spectrum of the conformal weights follows from the integrability
requirements:
\begin{equation}
              h=\frac{1}{2} (1+m) +i \nu \;\;\;\;\; {\rm and}\;\;\;\;\;
\overline{h}=\frac{1}{2} (1-m) +i \nu \;,
\label{hspectr}
\end{equation}
where $m$ is an integer. However, as can be seen from the general form
of the wave function (\ref{Psi}), $m$ has to be a multiple of 3 in order to
make $\Phi(z, \overline{z})$ single-valued.
Singlevaluedness of $\Phi $ alone allows for another condition:
\begin{equation}
              h=\overline{h}=\frac{1}{2} (1+\mu) +i \nu \;,
\label{hspectr2}
\end{equation}
where $\mu$ is an arbitrary real number. This condition, however, does not
correspond to the physical odderon state.

The single-valued wave function can be constructed only if both sectors,
holomorphic and antiholomorphic, are considered. For example in the vicinity
of $\xi =1$ and $\overline{\xi }=1$ the wave function of the whole system
reads:

\begin{equation}
\Phi _{h\overline{h}\,q\overline{q}}^{(1)}
(z,\xi, \overline{z}, \overline{\xi })=
z^{h/3} \; \overline{z}^{\overline{h}/3}  \;
{\vec{v}%
^{(1)\;{\rm T}}}(\overline{\xi };\overline{q})\,A^{(1)}(\overline{q},q)\,%
\vec{u}^{(1)}(\xi ;q)\;.  \label{Psi1}
\end{equation}
Similarly in the vicinity of $\xi =-1$ and $\overline{\xi }=-1$ we have:
\begin{equation}
\Phi _{h\overline{h}\,q\overline{q}}^{(-1)}
(z,\xi, \overline{z}, \overline{\xi })=
z^{h/3} \; \overline{z}^{\overline{h}/3} \; {\vec{v}%
^{(-1)\;{\rm T}}}(\overline{\xi };\overline{q})\,A^{(-1)}(\overline{q},q)\,%
\vec{u}^{(-1)}(\xi ;q).  \label{Psi-1}
\end{equation}

The requirement that the wave function $\Phi $ should be single-valued,
leads to the observation that matrices $A^{(\pm 1)}$ have to be diagonal:
\begin{equation}
A^{(-1)}=\left[
\begin{array}{lll}
\alpha  & 0 & 0 \\
0 & \beta  & 0 \\
0 & 0 & \gamma
\end{array}
\right] \;,\;\;\;\;\;\;A^{(1)}=\left[
\begin{array}{lll}
\alpha ^{\prime } & 0 & 0 \\
0 & \beta ^{\prime } & 0 \\
0 & 0 & \gamma ^{\prime }
\end{array}
\right] \;.  \label{matrAm}
\end{equation}
However, because the two solutions (\ref{Psi1},\ref{Psi-1}) are related by
Eq.(\ref{ancont}), we get the following relation:
\begin{equation}
\overline{\Gamma }^{\text{T}}(\overline{q})\,A^{(-1)}(\overline{q}%
,q)\,\Gamma (q)=A^{(1)}(\overline{q},q)\,.  \label{cond1}
\end{equation}
Introducing $\overrightarrow{a}=(\alpha ,\beta ,\gamma )$ and $%
\overrightarrow{b}=(\alpha ^{\prime },\beta ^{\prime },\gamma ^{\prime })$
we can conveniently rewrite Eq.(\ref{cond1}) in the matrix form:
\begin{equation}
C_{{\rm up}}\,\overrightarrow{a}=0\ \;,\qquad C_{{\rm low}}\ \overrightarrow{%
a}=0\;\;\;\;\text{ and\qquad \ }B\,\overrightarrow{a}=\,\overrightarrow{b}\,,
\label{cond2}
\end{equation}
where matrix $C_{{\rm up}}$, corresponding to the 3 zeros above the diagonal
of matrix $A^{(1)}$, takes the following form:
\begin{equation}
C_{{\rm up}}=\left[
\begin{array}{lll}
\overline{\Gamma }_{11}\Gamma _{12} & \overline{\Gamma }_{21}\Gamma _{22} &
\overline{\Gamma }_{31}\Gamma _{32} \\
\overline{\Gamma }_{11}\Gamma _{13} & \overline{\Gamma }_{21}\Gamma _{23} &
\overline{\Gamma }_{31}\Gamma _{33} \\
\overline{\Gamma }_{12}\Gamma _{13} & \overline{\Gamma }_{22}\Gamma _{23} &
\overline{\Gamma }_{32}\Gamma _{33}
\end{array}
\right]   \label{matrC}
\end{equation}
and matrix $C_{{\rm low}}$ , corresponding to the zeros below the diagonal
of $A^{(1)}$, is obtained from $C_{{\rm up}}$ by interchanging $\overline{%
\Gamma }\leftrightarrow \Gamma $. Finally matrix $B$ reads:
\begin{equation}
B=\left[
\begin{array}{lll}
\overline{\Gamma }_{11}\Gamma _{11} & \overline{\Gamma }_{21}\Gamma _{21} &
\overline{\Gamma }_{31}\Gamma _{31} \\
\overline{\Gamma }_{12}\Gamma _{12} & \overline{\Gamma }_{22}\Gamma _{22} &
\overline{\Gamma }_{32}\Gamma _{32} \\
\overline{\Gamma }_{13}\Gamma _{13} & \overline{\Gamma }_{23}\Gamma _{23} &
\overline{\Gamma }_{33}\Gamma _{33}
\end{array}
\right] \,.  \label{matrB}
\end{equation}

Quantization conditions follow from the requirement that there exist
non-zero solutions of Eq.(\ref{cond2}) for $\alpha $, $\beta $ and $\gamma $:
\begin{equation}
{\rm Det}C_{{\rm up}}=0\;,\;\;\;\; {\rm and}\;\;\;\;
{\rm Det}C_{{\rm low}}=0\;.
%\;\;{\rm and}\;\;\;{\rm Det}B\ne 0\;.
\label{cond3}
\end{equation}
Moreover the first two equations in (\ref{cond2}) should be uniquely
solvable for $\alpha $, $\beta $ and $\gamma $ in function of one free
parameter (which is not automatic even if Eqs.(\ref{cond3}) are satisfied).

Let us now discuss numerical solutions of the quantization conditions (\ref
{cond3}). We have found that the zero eigenvalue of the
odderon charge exists always for arbitrary conformal weight. Korchemsky \cite
{KorBethe} has argued that these states should be excluded from the spectrum
of the odderon Hamiltonian (\ref{ham}), since it is not clear if they are
normalizable (see however Ref.\cite{bootstrap}). In what follows we shall
concentrate only on the imaginary solutions for $q$  which, as already said,
are relevant for the odderon system.

We have first tried to look for solutions of the quantization
conditions (\ref{cond3}) corresponding to the spectrum of $h$ and
$\overline{h}$ given by Eq.(\ref{hspectr}) with $\nu=0$.  The lowest
two values of $m$ for which non-zero, imaginary solutions for $q$ exist
are equal  $m =0$ and $\mid m \mid =3$.  There also exist solutions for
higher $\mid m \mid $, which will not be discussed in this note.  For
the above values of $m$, with our choice of phases in Eq.(\ref{sols}),
both Det$C_{{\rm up}}$ and Det$C_{{\rm low}}$  are imaginary along the
imaginary axis in the complex $q$ plane. Moreover, for imaginary $q$:
Det$C_{{\rm up}}$=Det$C_{{\rm low}}\equiv $ Det$C$.  In Fig.1 we plot
Im$\;$Det$C$ as a function of Im$\;q$ for $h=1/2$, or equivalently
$m=0$ (solid line)  and for $\mid m \mid=3 $ (dashed line).  We see
that Im$\;$Det$C$ is an antisymmetric, oscillating function of Im$\;q$,
with amplitude growing with $ \mid {\rm Im}\;q\mid $. Zeros of the
functions in Fig.1 correspond to the quantized values of $q$. For $m=0$
two non-zero eigenvalues are visible: $q=\pm 0.2052575\times i$ and
$q=\pm 2.34392\times i$.  We have found two more eigenvalues of
Im$\;q$: $\pm 8.326346$ and $\pm 20.080497$.  For higher $q$'s care
must be taken in order not to loose numerical stability.  These
eigenvalues have been found previously in Ref.\cite{JaWoOddsol}.  For
$\mid m \mid =3$ only one non-zero eigenvalue of $q$ is visible in
Fig.1, namely $\pm 1.176667 \times i$.  We have also found two next
eigenvalues corresponding to Im$\;q =\pm 6.35591$ and $\pm 17.69346$.

Next, let us consider condition (\ref{hspectr2}) also with $\nu=0$.
Here solutions are found for arbitrary real $\mu$. In Fig.2 we plot Im$\;$%
Det$C$ as a function of Im$\;q$ for $h=1/2$ (solid line) and for $h=0$ and $1
$ (both denoted by dashed line).
As soon as we move $h$ away from 1/2 the little wiggle, seen for 
$h=1/2$ (or equivalently $\mu=0$),
straightens up and the two eigenvalues symmetrically drift towards zero.
Eventually, for $h$=0
and for $h$=1, they reach zero value. This is depicted in Fig.3 where the
drift of the first 2 positive eigenvalues of $\hat{q}$ is plotted in
dependence on $h$. Second eigenvalue reaches zero for $h=3$ and $h=-2$.
This kind of behavior is observed for all imaginary eigenvalues, for
which our numerical procedures are stable.

It is also interesting to consider complex $h=1/2\pm i\,\nu $.
As soon as one varies $\nu$, $i\,q$ grows up as $\nu $ increases
\cite{newBraun}. Our results agree with the ones of Ref.\cite{newBraun}.

As a cross-check on our method we have also solved
quantization conditions for $h=3$, 4, 5 and 6. For these values of $h$
Korchemsky in Ref.\cite{KorBethe} found real spectrum of $\hat{q}$ 
for the polynomial solutions of the pertinent Baxter equations. Our
quantization conditions are more general, so we find more eigenvelues
for integer $h$, among them the ones reported in Ref.\cite{KorBethe}.
It is interesting to note, that for these values of $h$, there exist
real solutions corresponding both to Eq.(\ref{hspectr}) and
Eq.(\ref{hspectr2}), and moreover these solutions are equal.

To summarize: in this short note we have solved quantization conditions
for the odderon charge $\hat{q}$ which has been identified with the
cross-product of three conformal spins. We have proposed to study the
odderon equation in terms of a new variable called $\xi $, which was
earlier discussed by Janik in the context of the modular invariance of
the odderon system \cite{Janik}. We have solved quantization conditions
for the odderon charge using the method recently proposed by Janik and
Wosiek \cite {JaWoOddsol}. Our approach can be straightforwardly
applied for any conformal weight $h$. For $h=1/2$ we have reproduced
eigenvalues found in Ref.\cite{JaWoOddsol}. For integer $h\ge 3$ we
have reproduced eigenvalues of $\hat{q}$ found by Korchemsky in
Ref.\cite{KorBethe}. To illustrate the possibility of solving the
quantization conditions for arbitrary $h$ we have studied the drift of
the lowest eigenvalues of $\hat{q}$ for real $h$ in the vicinity of
$h=1/2$ using two different quantization conditions for $h$
(\ref{hspectr},\ref{hspectr2}). For the physical odderon state
(\ref{hspectr}) the lowest values of $h$ for which Eq,(\ref{cond3})
could have been found correpond to $m=0$ and $\pm 3$. For the
unphysical condition  $h=\overline{h}$  continous sets of solutions
exist for real $h$.  Further results, also for complex $h$ and for real
values of $q$, will be discussed in the forthcoming paper
\cite{PraRos2}.

\section*{ACKNOWLEDGEMENTS}

We thank R. Janik and J. Wosiek for discussions. Special thanks for
hospitality are due to K. Goeke and the Institute of Theoretical Physics II
of the Ruhr-University in Bochum, Germany, where most of this work was done.
Partial support of Polish KBN Grant PB 2 PO3B 044 12 is acknowledged. M.P.
acknowledges support of Alexander von Humboldt Stiftung.

%%%%%%%%%%%%%FIGURES
\newpage

\begin{figure}[tbp]
\hspace{4cm}\epsfbox{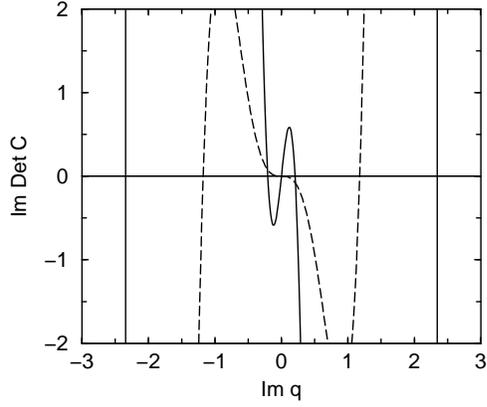}
\caption{Im~Det$\; C$ as function of Im$\; q$ for $m=0$ (solid line) and
for $\mid m \mid =3$ (dashed curve).}
\end{figure}
\begin{figure}[tbp]
\hspace{4cm}\epsfbox{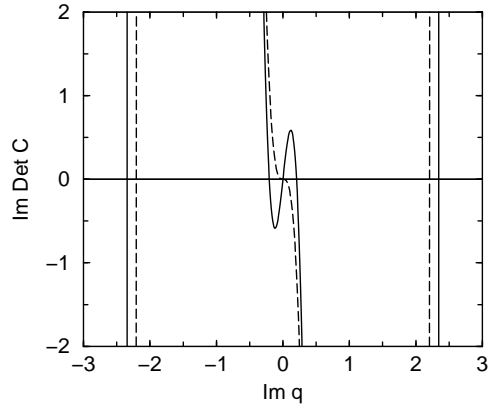}
\caption{The same as FIG.1 but for $h=\overline{h}$. Solid curve corresponds
to $h=1/2$, whereas dashed line corresponds both to $h=0$ and $h=1$.}
\end{figure}
\begin{figure}[tbp]
\hspace{4cm}\epsfbox{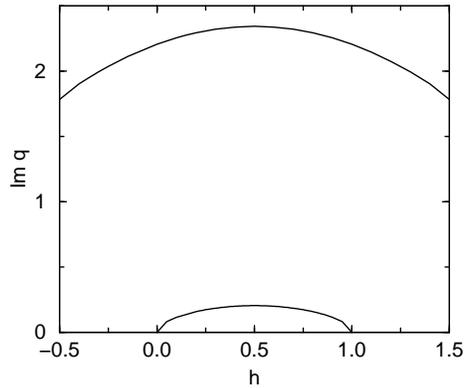}
\caption{Positive part of Im$\; q$ of two first eigenvalues as functions of
(real) $h=\overline{h}$.}
\end{figure}

%%%%%%%%%%%%%%REFERENCES


\begin{references}
\bibitem{KLF}  E.A. Kuraev, L.N.Lipatov and V.S.Fadin, Zh. Eksp. Teor. Fiz.
{\bf 71} (1976) 840 (JETP {\bf 44} (1976) 433, {\bf 45} (1977) 199).

\bibitem{BL}  Ya.Ya.Balitzky and L.N.Lipatov, Yad. Fiz. {\bf 28} (1978) 1597
(Sov. J. of Nucl. Phys. {\bf 28} (1978) 822).

\bibitem{Bart80}  J. Bartels, Nucl. Phys. {\bf B175} (1980) 365.

\bibitem{KwPra}  J.Kwieci{\'{n}}ski and M.Prasza{\l }owicz, Phys. Lett. {\bf %
B94} (1980) 413.

\bibitem{Jar}  T.Jaroszewicz, {\em High-energy multi-gluon ex{\-}change
amp{\-}li{\-}tudes}, Trieste pre{\-}print IC/80/175, see also Acta Phys.
Pol. {\bf B11} (1980) 965.

\bibitem{Lipatov}  L.N.~Lipatov, Phys.Lett. {\bf B251} (1990) 284; {\bf B309}
(1993) 394.

\bibitem{Lipsolv}  L.N. Lipatov, Pisma. Zh. Eksp. Teor. Fiz. {\bf 59} (1994)
569 (JETP Lett. {\bf 59} (1994) 571).

\bibitem{FadKor}  L.D. Faddeev and G.P. Korchemsky, Phys. Lett. {\bf B342}
(1995) 311.

\bibitem{KorBethe}  G.P. Korchemsky, Nucl. Phys. {\bf B443} (1995) 255.

\bibitem{GLN}  P.Gauron, L.Lipatov and B.Nicolescu, Phys. Lett. {\bf B260}
(1991) 407; Phys. Lett. {\bf B304} (1993) 334; Z. Phys. {\bf C63} (1994) 253
.

\bibitem{ArmBraun}  N. Armesto and M.A. Braun, {\em The intercept of
symmetric multi--gluon configurations in the variational approach}, {\tt %
hep-ph/9410411}, Santiago de Compostela preprint US-FT-9-94.

\bibitem{bootstrap}  N. Armesto and M.A.Braun, Z. Phys. {\bf C75} (1997) 709.

\bibitem{Braun98}  M.A.Braun, {\em On the odderon intercept in the
variational approach}, {\tt hep-ph/9801352}, St.Petersburg preprint
SPbU-IP-1998/3.

\bibitem{Korch}  G.P. Korchemsky, Nucl. Phys. {\bf B462} (1996) 333; {\em %
WKB quantization of reggeon compound states in high-energy QCD}, {\tt %
hep-ph/9801377}, Orsay preprint LPTHE-97-73.

\bibitem{JaWoOdd}  R. Janik and J. Wosiek, Phys.Rev.Lett. {\bf 79} (1997)
2935.

\bibitem{JaWoOddsol}  R. Janik and J. Wosiek, {\em Solution of the odderon
problem}, {\tt hep-ph/9802100}, Cracow preprint TPJU-2/98.

\bibitem{Janik}  R. Janik, Phys.Lett. {\bf B371} (1996) 293; Acta
Phys.Polon. {\bf B27} (1996) 1275.


\bibitem{newBraun}  M.A.Braun, {\em On the odderon intercept in the
perturbative QCD}, {\tt hep-ph/9804432}, St.Petersburg preprint
SPbU-IP-1998/8.

\bibitem{PraRos2}  M.~Prasza{\l}owicz and A.~Rostworowski in preparation.
\end{references}
\end{document}